\documentclass{ocephys}

\newcommand{\pp}[2]{\frac{\partial #1}{\partial #2}}

\title{Closure of the sea surface height budget \\ with a Stokes offset}

\author[a]{J\"orn Callies}
\author[a,b]{Charly de Marez}
\author[c]{Jinbo Wang}
\author[d]{Bruce Haines}
\affil[a]{California Institute of Technology}
\affil[b]{University of Iceland}
\affil[c]{Texas A\&M University}
\affil[d]{Jet Propulsion Laboratory, California Institute of Technology}

\corr{J\"orn Callies, \href{mailto:jcallies@caltech.edu}{jcallies@caltech.edu}}
\disclaimer{This work has been accepted to the Journal of Physical Oceanography. The American Meteorological Society does not guarantee that the copy provided here is an accurate copy of the version of record.}

\begin{document}

\nolinenumbers

\section*{Abstract}

The sea surface height budget, obtained by integrating hydrostatic balance over the water column, relates sea surface height variations to variations of the seafloor pressure, density in the water column, and atmospheric surface pressure. This budget is crucial for calibrating and interpreting satellite altimetry measurements. It only holds once non-hydrostatic surface gravity waves are averaged out, however, which complicates an observational closure of the budget. Using data from the California Current System, this study demonstrates that the budget closes to within understood uncertainties if GPS buoy measurements of surface height are interpreted as Lagrangian measurements. The buoy largely follows wave motion and spends slightly more time near wave crests than troughs. The associated Stokes offset, which reaches a maximum of \qty{16}{\centi\meter} in these observations, must be accounted for in the Eulerian sea surface height budget.

\section{Introduction}

Precise measurements of the sea surface height from space have greatly advanced global oceanography. Satellite altimetry captures the geostrophic circulation, which consists of gyres, boundary currents, jets, and ubiquitous mesoscale eddies \parencite{wunsch_satellite_1998,chelton_global_2011}. Altimetry data are the basis of accurate maps of the external tides \parencite{le_provost_ocean_1995}; they have revealed the long-range propagation of internal tides \parencite{ray_surface_1996}; and they can be used to infer where the external tides lose energy to internal tides and dissipation \parencite{egbert_significant_2000}, with important implications for the circulation and stratification of the deep ocean \parencite{munk_abyssal_1998}. Furthermore, altimetry data constitute the backbone of estimates of global sea level rise \parencite{cazenave_contemporary_2010}.

The Surface Water and Ocean Topography (SWOT) mission, launched in December 2022, measures the sea surface height with substantially higher precision than conventional nadir altimetry \parencite{fu_surface_2024}. To evaluate these new measurements, a set of moorings was deployed in the California Current System \parencite{wang_swot_2025}. These moorings measure density fluctuations in the water column, the vertical integral of which dominates the non-tidal sea surface signal \parencite{wang_development_2022}. Some of these moorings also carry GPS buoys that directly measure the sea surface height using recently developed technology \parencite{haines_calval_2017,desai_status_2018,haines_harvest_2019}, adding redundancy to the evaluation of SWOT data. A comparison between these GPS buoy data and altimetric measurements should be made with caution, however, because the buoy measurements are subject to a Stokes offset due to surface gravity waves \parencite{longuet-higgins_eulerian_1986}. The buoys move mainly with the gravity wave orbit, the trajectory of a fluid parcel in the wave field, spending slightly more time near the wave crest than near the wave trough, shifting the Lagrangian mean position upward relative to the Eulerian mean sea surface height. Using data from a pre-launch deployment, we demonstrate the importance of this Stokes offset by closing the sea surface height budget, which is obtained by integrating the hydrostatic balance over the water column \parencite{gill_theory_1973}.

\section{The sea surface height budget}

At sufficiently large horizontal scales, ocean flows have a small aspect ratio and nearly satisfy hydrostatic balance,
\begin{equation}
  \pp{p}{z} = -\rho g,
  \label{eqn:hydrostatic}
\end{equation}
where $p$ is pressure, $\rho$ is density, $g$ is the gravitational acceleration, and $z$ is the vertical coordinate referenced to the long-term Eulerian mean position of the sea surface. Integrating this balance from the seafloor at $z = -H$ to the surface at $z = h$ and denoting the bottom and surface pressures as $p_\mathrm{b}$ and $p_\mathrm{a}$, respectively, we obtain
\begin{equation}
  p_\mathrm{b} - p_\mathrm{a} = g \int_{-H}^h \rho \, \d z \simeq \rho_0 g h + g \int_{-H}^0 \rho \, \d z.
  \label{eqn:hydint}
\end{equation}
In the approximation above, we use the fact that surface density variations away from the reference $\rho_0 = \qty{1024}{\kilo\gram\per\meter\cubed}$ are small compared to the reference value and that the sea surface height variations~$h$ are small compared to the reference water depth~$H$. This integrated hydrostatic balance can be written as the sea surface height budget
\begin{equation}
  h - h_\mathrm{a} - h_\mathrm{b} = h_\mathrm{s} - H,
  \label{eqn:sshbudgetfull}
\end{equation}
where
\begin{equation}
  h_\mathrm{a} = -\frac{p_\mathrm{a}}{\rho_0 g}, \qquad h_\mathrm{b} = \frac{p_\mathrm{b}}{\rho_0 g}, \qquad h_\mathrm{s} = -\int_{-H}^0 \frac{\rho - \rho_0}{\rho_0} \, \d z.
\end{equation}
The terms in the sea surface height budget are the surface height~$h$ itself, the barometric correction~$h_\mathrm{a}$ \parencite[also known as the ``inverted barometer'' effect;][]{wunsch_atmospheric_1997}, the height equivalent to the bottom pressure~$h_\mathrm{b}$, the steric height~$h_\mathrm{s}$, and the mean water depth~$H$. If we remove the time mean of all quantities, we obtain
\begin{equation}
  h' - h'_\mathrm{a} - h'_\mathrm{b} = h'_\mathrm{s}.
  \label{eqn:sshbudget}
\end{equation}
Closure of the budget~\eqref{eqn:sshbudget} to within understood uncertainties would increase our confidence in the measurements used to estimate each term. It would also increase our confidence in estimates of any one term as a residual of this budget, for example, estimates of the sea surface height from a combination of measurements of bottom pressure and steric height, supplemented with an estimate of atmospheric pressure from a surface measurement or reanalysis product.

\begin{figure}[t]
  \centering
  \includegraphics[scale=0.6]{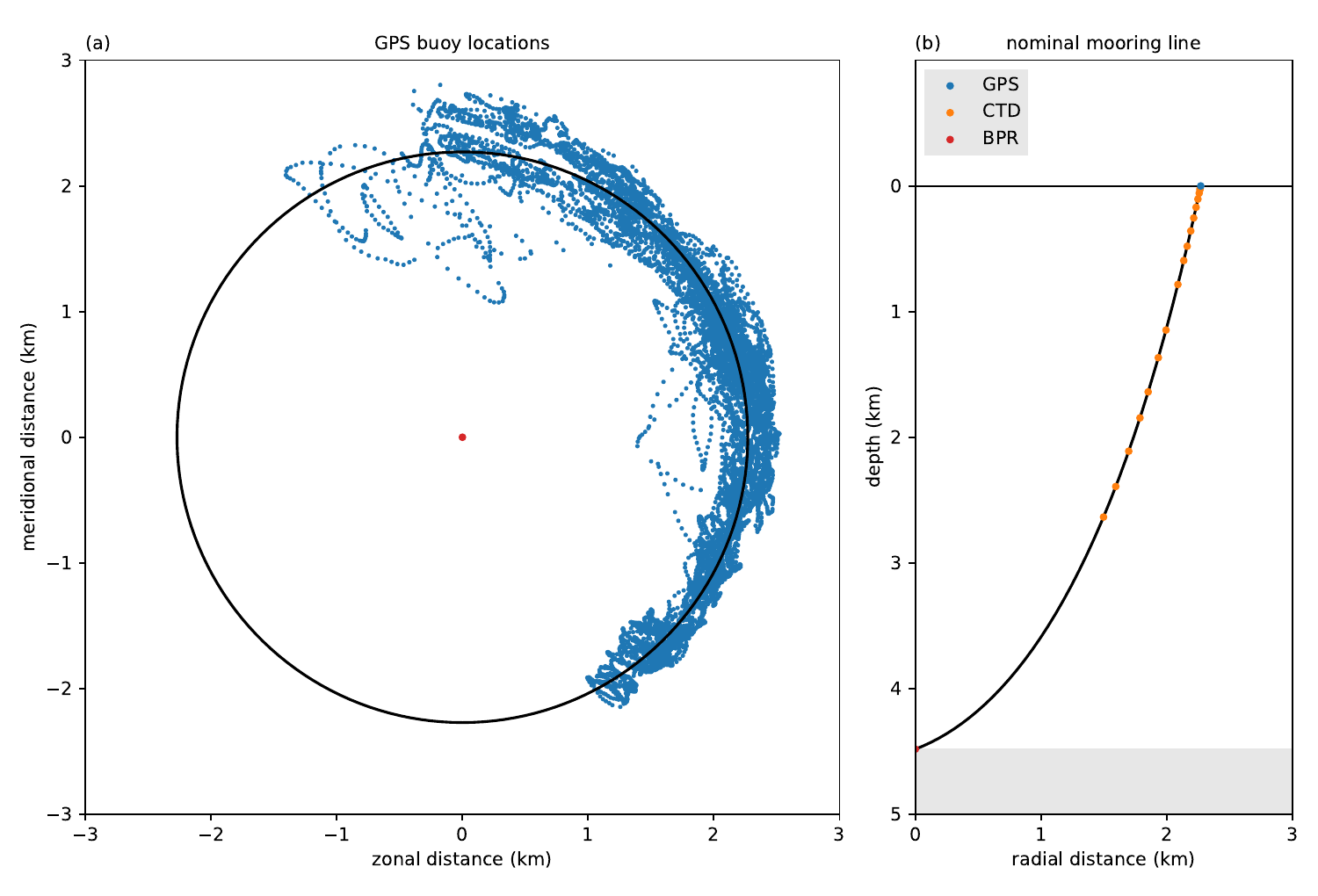}
  \caption{Configuration and position of the mooring in the California Current System. (a)~Location of the GPS buoy over the course of the deployment (blue dots), relative to the mooring anchor (red dot, \ang{36}07' N, \ang{125}08' E), inferred by fitting a circle to the measured GPS surface positions (black circle). (b)~Nominal mooring line and sensor positions when the surface buoy is at its mean displacement (black circle).}
  \label{fig:mooring}
\end{figure}

We demonstrate here that the budget~\eqref{eqn:sshbudget} can be closed using mooring measurements from a field experiment in the California Current System that was conducted from September~2019 to January~2020 in anticipation of the SWOT launch \parencite{wang_development_2022}. We use measurements of the sea surface height from a GPS buoy, estimate the barometric correction using atmospheric reanalysis \parencite[ERA5 hourly sea level pressure;][]{hersbach_era5_2020}, use a bottom pressure recorder, and estimate steric height from a set of 17~CTDs along the mooring line (Fig.~\ref{fig:mooring}). We remove the first two weeks of the bottom pressure data because of instrument drift during that period. The mooring was designed as a slack mooring, with the mooring line much longer than the water depth (\qty{5190}{\meter} vs.\ \qty{4480}{\meter}), which reduces the line tension and allows the surface buoy to follow the surface height. A detailed description of the mooring design and the instrumentation, including a discussion of error sources, is provided by \textcite{wang_development_2022}; the mooring used here is the ``northern mooring'' described there. The only difference in the processing of the GPS data is that we do not apply the empirical correction for an apparent sea state bias, which was estimated as $0.018 \eta$ in \textcite{wang_development_2022}, where $\eta$ is the significant wave height. We argue below that this apparent bias is due principally to the Stokes offset. For all time series, we remove a linear fit over the period during which all observations are available (2019-09-19, 15:00 UTC to 2020-01-19, 15:00 UTC). These removed trends are much smaller than the dominant signal for all measurements; we remove them to avoid remaining instrument drift to affect the analysis.

\begin{figure}[t]
  \centering
  \includegraphics[scale=0.6]{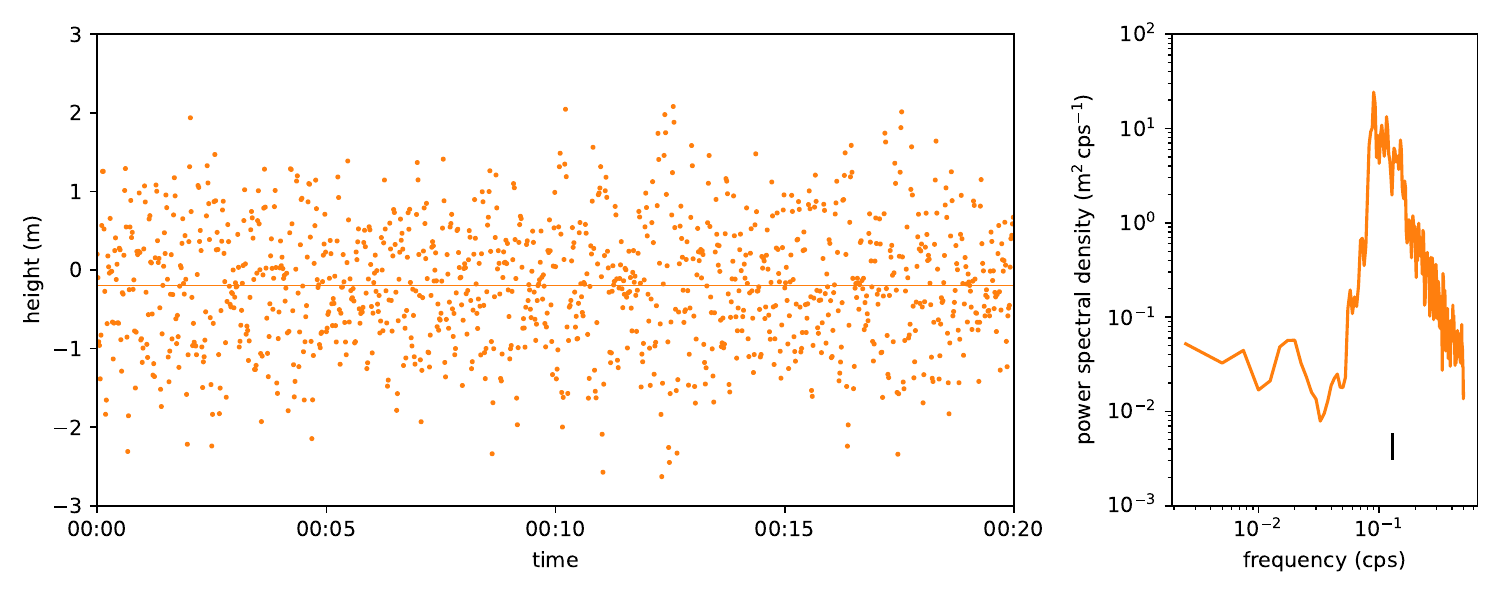}
  \caption{Example surface height signal from the GPS buoy dominated by surface gravity waves. (a)~Time series of the barometrically corrected \qty{1}{\hertz} GPS buoy data for the first 20~minutes of 2019-10-11 (UTC), with the mean over this time period~$h_\mathrm{g}' - h_\mathrm{a}' = \qty{-20}{\centi\meter}$ (relative to the time series mean) indicated by the horizontal line. (b)~Variance spectrum estimated from this time series showing a peak corresponding to the dominant surface gravity waves. The dominant wave period of~\qty{7.7}{\second}, calculated from the first mode of this spectrum, is indicated by the black vertical line.}
  \label{fig:waveexample}
\end{figure}

As expected, the dominant signal on the GPS buoy is that of wind waves and swell (Fig.~\ref{fig:waveexample}). These are surface gravity waves in the deep-water limit and do not satisfy~\eqref{eqn:hydrostatic}. We must average over surface gravity waves to close the sea surface height budget for the signals produced by larger-scale hydrostatic flows. We average here over 20-minute bins. Because the buoy largely moves with the surface waves in both the vertical and horizontal directions, a time average over the buoy measurements is closer to a Lagrangian than an Eulerian mean of the sea surface height. For a plane wave and to second order in the wave amplitude~$a$, this Lagrangian mean differs from the Eulerian mean surface height by the Stokes offset $h_\mathrm{l} = \frac{1}{2} k a^2$, where $k$ is the wavenumber \parencite[e.g.,][]{longuet-higgins_eulerian_1986,buhler_waves_2014,salmon_more_2020}.\footnote{The Stokes offset is linked to the divergent generalized Lagrangian-mean (GLM) circulation when the wave field grows or decays in amplitude \parencite{mcintyre_note_1988,buhler_waves_2014,vanneste_stokes_2022}.} The Eulerian mean surface height~$h$ and the Lagrangian mean~$h_\mathrm{g}$ measured by the GPS buoy are therefore related by
\begin{equation}
  h = h_\mathrm{g} - h_\mathrm{l}.
\end{equation}
For a typical significant wave height of $\eta = \qty{3}{\meter}$ and a typical wavelength of $2\pi/k = \qty{100}{\meter}$, the Stokes offset is $h_\mathrm{l} = \frac{1}{16} k \eta^2 = \qty{3.5}{\centi\meter}$, where we used that the significant wave height is defined as four times the root-mean-square height. The offset is therefore comparable to the mesoscale signal (order~\qty{10}{\centi\meter}) and greater than SWOT's precision (order~\qty{1}{\centi\meter}). It varies in time as the surface wave properties evolve, so it is indispensable to close the budget~\eqref{eqn:sshbudget}. The residual $h_\mathrm{g}' - h_\mathrm{a}' - h_\mathrm{b}' - h_\mathrm{s}'$ should not vanish but equal the Stokes offset~$h_\mathrm{l}'$ (with its average over the time series removed).

\begin{figure}[tp]
  \centering
  \includegraphics[scale=0.6]{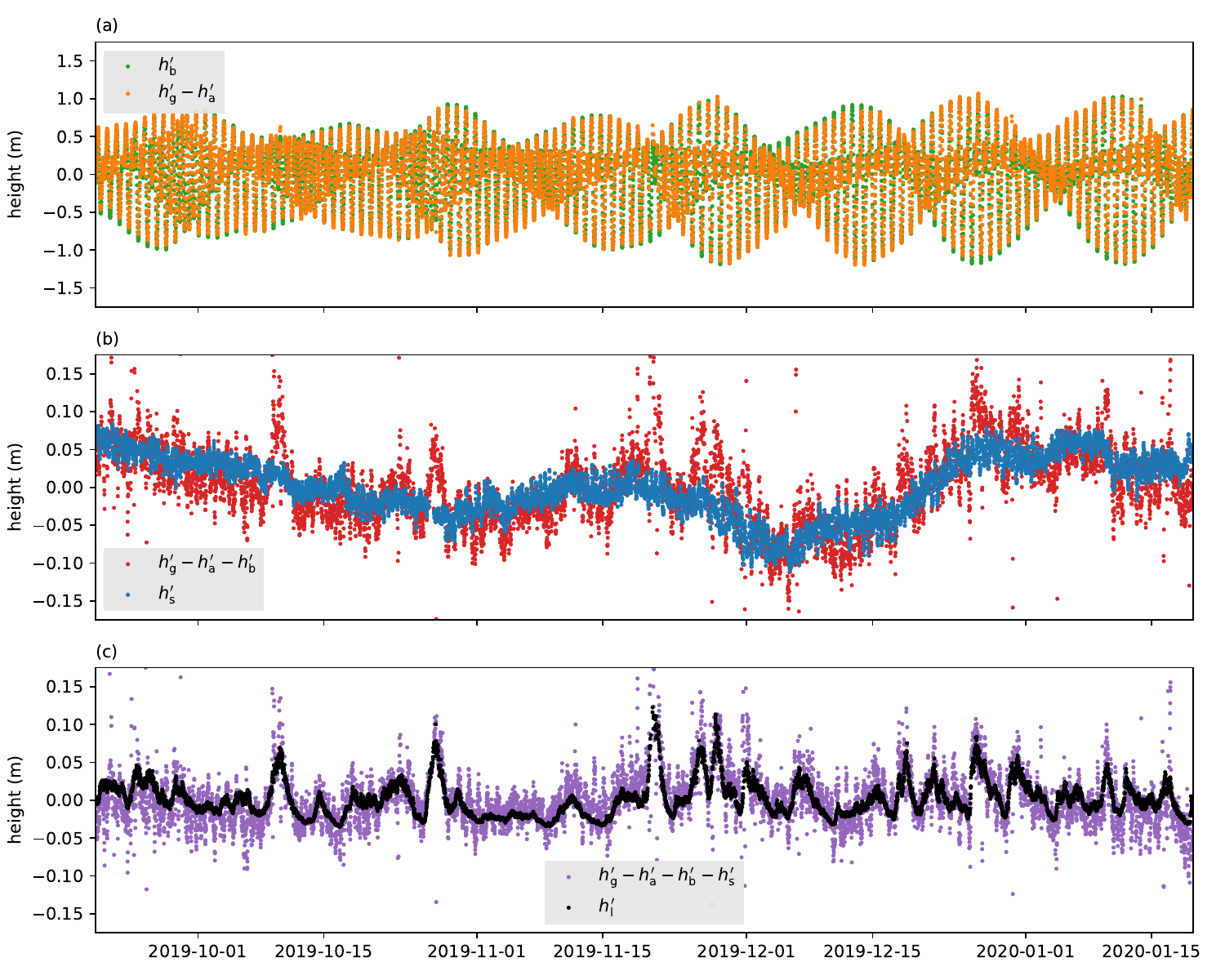}
  \caption{Closure of the sea surface height budget by the Stokes offset. (a)~The height equivalent to the bottom pressure~$h_\mathrm{b}'$ inferred from the bottom pressure recorder and the barometrically corrected surface height~$h_\mathrm{g}' - h_\mathrm{a}'$ inferred from the GPS buoy and corrected using atmospheric reanalysis. (b)~Residual~$h_\mathrm{g}' - h_\mathrm{a}' - h_\mathrm{b}'$ and the steric height~$h_\mathrm{s}'$ inferred from CTDs along the mooring line. (c)~Residual of the sea surface height budget $h_\mathrm{g}' - h_\mathrm{a}' - h_\mathrm{b}' - h_\mathrm{s}'$ and the Stokes offset~$h_\mathrm{l}'$ calculated from the wave properties estimated from the GPS buoy according to~\eqref{eqn:stokes} and with a time mean over the observational period removed.}
  \label{fig:heights}
\end{figure}

Once the surface gravity waves are averaged out, the barometrically corrected surface height $h_\mathrm{g}' - h_\mathrm{a}'$ is dominated by the external tides with an amplitude on the order of \qty{1}{\meter} (Fig.~\ref{fig:heights}a). As expected, this tidal signal is largely mirrored in the bottom pressure. The residual $h_\mathrm{g}' - h_\mathrm{a}' - h_\mathrm{b}'$ is reduced in amplitude to about \qty{10}{\centi\meter} and exhibits prominent variations on a time scale of a month or two (Fig.~\ref{fig:heights}b). This signal is due to mesoscale eddies moving across the mooring \parencite{wang_development_2022,de_marez_observational_2023}. It is also captured by the steric height~$h_\mathrm{s}'$, as expected from the sea surface height budget~\eqref{eqn:sshbudget}. The time series $h_\mathrm{g}' - h_\mathrm{a}' - h_\mathrm{b}'$ has more variance than~$h_\mathrm{s}'$ at high frequencies and a number of outliers, which we make no attempt to remove. Nevertheless, it is clear that there are systematic differences between the two in the form of spikes in $h_\mathrm{g}' - h_\mathrm{a}' - h_\mathrm{b}'$ that last a few days and are absent in~$h_\mathrm{s}'$. These spikes are prominent in the residual of the sea surface height budget $h_\mathrm{g}' - h_\mathrm{a}' - h_\mathrm{b}' - h_\mathrm{s}'$ (Fig.~\ref{fig:heights}c). They coincide with periods of large-amplitude surface gravity waves, when the Stokes offset is large.

\begin{figure}[tp]
  \centering
  \includegraphics[scale=0.6]{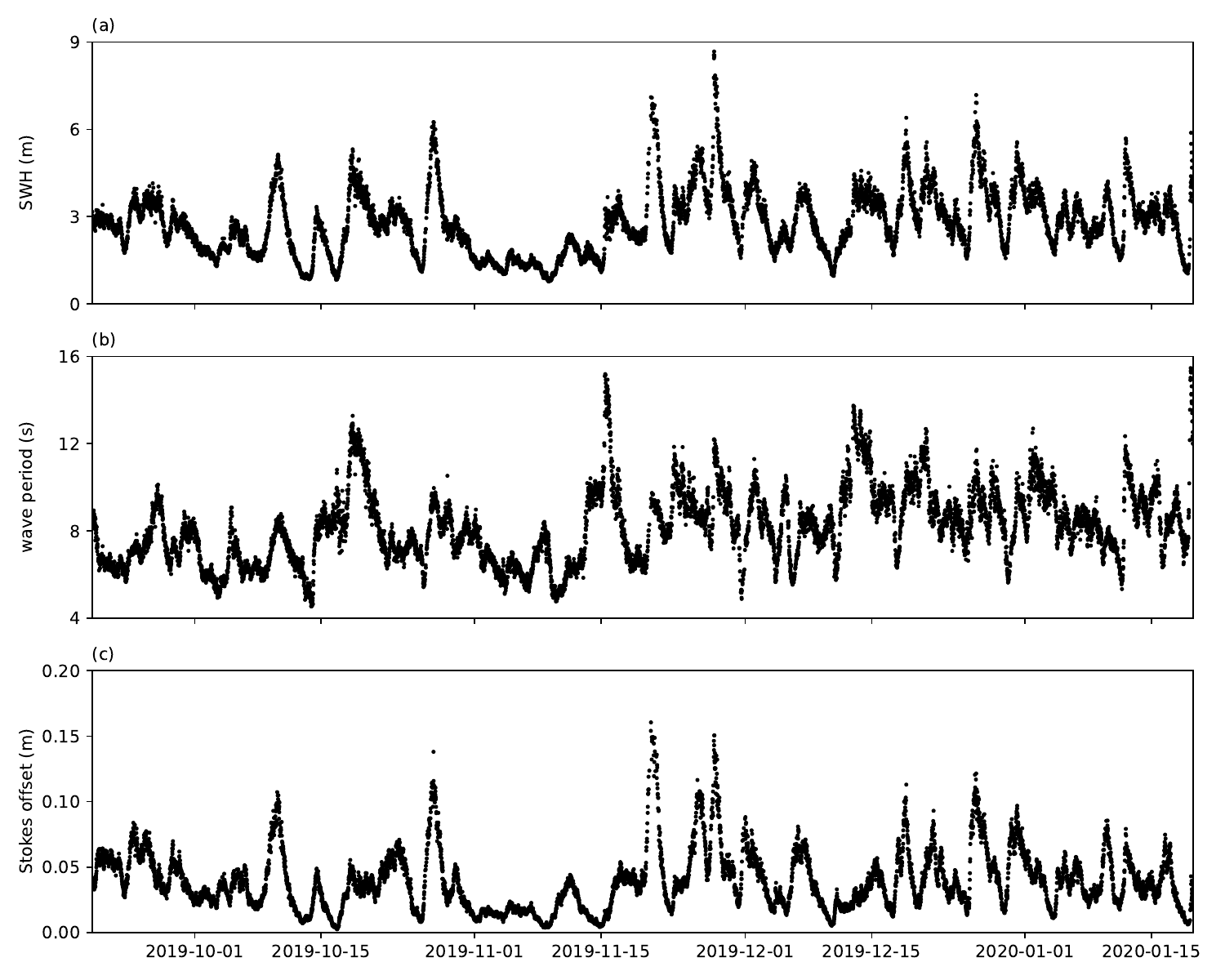}
  \caption{Wave properties and the Stokes offset. (a)~Significant wave height~$\eta$ calculated as four times the standard deviation of the GPS buoy measurements. (b)~Dominant wave period calculated from the first mode of the variance spectrum of the surface height measured by the GPS buoy. (c)~Stokes offset~$h_\mathrm{l}$ estimated by applying~\eqref{eqn:stokes} to the GPS buoy measurements.}
  \label{fig:waves}
\end{figure}

We estimate the Stokes offset from the observed wave properties. For each 20-minute averaging period, we estimate the variance spectrum~$S(\omega)$ from the barometrically corrected GPS buoy surface height (Fig.~\ref{fig:waveexample}). We chop each 20-minute segment into five chunks that overlap by 50\%, and we apply a Hann window to each chunk. The variance spectrum is estimated as the average over the spectrograms from the five chunks. From these wave spectra, we estimate the Stokes offset as, using the dispersion relation for deep-water waves $k = \omega^2/g$,
\begin{equation}
  h_\mathrm{l} = \int k S(\omega) \, \d \omega = \int \frac{\omega^2}{g} S(\omega) \, \d \omega.
  \label{eqn:stokes}
\end{equation}
This generalizes the formula for a single plane wave given above to a full spectrum of waves \parencite{kenyon_stokes_1969,webb_wave_2011}. The Stokes offset varies primarily because of variations in the wave amplitude and secondarily because of variations in the dominant wave period (Fig.~\ref{fig:waves}).

This Stokes offset~$h_\mathrm{l}'$, with its time mean removed, matches the residual of the sea surface height budget~$h_\mathrm{g}' - h_\mathrm{a}' - h_\mathrm{b}' - h_\mathrm{s}'$ remarkably well (Fig.~\ref{fig:heights}c). The peaks in the residual line up with periods of large-amplitude waves, when the Stokes offset is large. The height of the peaks matches well, and we emphasize that there is no adjustable parameter in this calculation. This demonstrates the importance of the Stokes offset in the closure of the sea surface height budget if the surface height is measured using a buoy that behaves largely like a Lagrangian particle in the surface gravity wave field. 

\begin{figure}[t]
  \centering
  \includegraphics[scale=0.6]{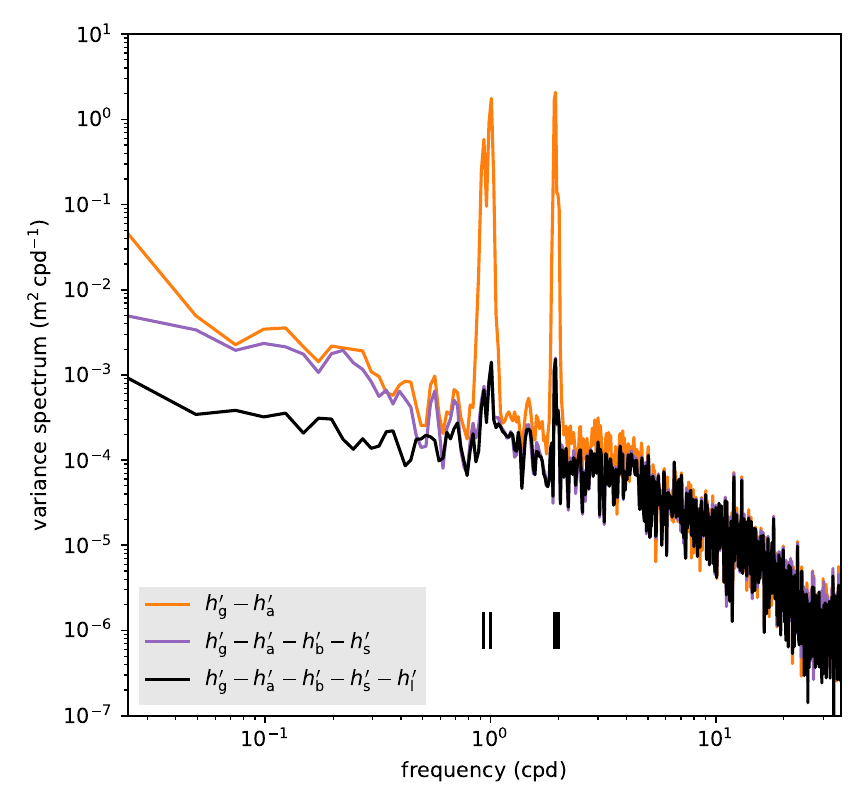}
  \caption{Variance spectra of the residual of the sea surface height budget with and without the Stokes offset ($h_\mathrm{g}' - h_\mathrm{a}' - h_\mathrm{b}' - h_\mathrm{s}' - h_\mathrm{l}'$ vs.\ $h_\mathrm{g}' - h_\mathrm{a}' - h_\mathrm{b}' - h_\mathrm{s}'$), as compared to the barometrically corrected surface height ($h_\mathrm{g}' - h_\mathrm{a}'$) from the GPS buoy. The frequencies of the O$_1$, K$_1$, M$_2$, and~S$_2$ tides are indicated by short black lines.}
  \label{fig:resspec}
\end{figure}

To quantify the reduction in the residual of the sea surface height budget~\eqref{eqn:sshbudget} when the Stokes offset is taken into account, we estimate frequency spectra of the residual with and without the Stokes offset ($h_\mathrm{g}' - h_\mathrm{a}' - h_\mathrm{b}' - h_\mathrm{s}' - h_\mathrm{l}'$ vs.\ $h_\mathrm{g}' - h_\mathrm{a}' - h_\mathrm{b}' - h_\mathrm{s}'$), and we compare these to the spectrum of the barometrically corrected height as measured by the GPS buoy ($h_\mathrm{g}' - h_\mathrm{a}'$). We fill gaps using linear interpolation and estimate the variance spectrum using the same procedure as described above, now applied to the time series of 20-minute averages. The spectra show that the Stokes offset is crucial for closing the sea surface height budget at periods greater than 1~day (Fig.~\ref{fig:resspec}). At these periods, the Stokes offset substantially reduces the residual and renders it small compared to $h_\mathrm{g}' - h_\mathrm{a}'$. At periods shorter than 1~day, the Stokes offset has little variance and makes no difference in the budget. The residual has variance comparable to $h_\mathrm{g}' - h_\mathrm{a}'$ at periods shorter than about 6~hours, indicating a failure to close the budget at these short periods. The residual also has significant peaks at the tidal frequencies, and the Stokes offset makes no substantial difference there. Overall, taking the Stokes offset into account reduces the root-mean-square residual from \qtyrange{3.6}{2.8}{\centi\meter}. We note that these numbers may be further reduced by outlier detection, longer averaging windows, and corrections for remaining biases in the GPS measurements; \textcite{wang_development_2022} found a \qty{2}{\centi\meter} root-mean-square residual in 1-hour averages when applying their empirical sea state bias correction.

\section{Discussion}

The empirical correction of an apparent sea state bias that \textcite{wang_development_2022} applied to the GPS buoy measurements is likely dominated by the Stokes offset. A linearization of the Stokes offset~$\frac{1}{16} k \eta^2$ at mean wave conditions yields a proportionality constant of $\frac{1}{8} k \eta = 0.021$, not far from the \num{.018} that \textcite{wang_development_2022} inferred from a linear fit to the GPS measurements against~$\eta$. That said, other sea state--dependent errors in the GPS solution, for example, changes in the effective elevation cutoff \parencite[cf.,][]{park_foliage_2010} from the varying wave field, cannot be dismissed and warrant further investigation.

There are signals in the residual of the sea surface height budget that are not explained by the Stokes offset, of course, most prominently at tidal periods and periods shorter than 6~hours (Figs.~\ref{fig:heights}c,~\ref{fig:resspec}). Clearly, errors persist in all measurements, and we refer the reader to \textcite{wang_development_2022} for an in-depth discussion. Errors in the GPS measurements due to uncompensated atmospheric refraction and errors in the correction for the mooring line tension as well as the incomplete coverage of the water column by CTDs (Fig.~\ref{fig:mooring}) are likely major contributors to the remaining residual. In addition, non-hydrostatic internal waves that do not satisfy~\eqref{eqn:hydrostatic} may contribute to the high-frequency residual. Overall, however, inclusion of the Stokes offset further strengthens the prospect of using GPS buoys as effective open-ocean tide gauges for the calibration and evaluation of the SWOT mission, as opposed to the CTD-based approach used so far \parencite{wang_swot_2025}.

In our calculation of the Stokes offset, we assumed that the GPS buoy perfectly follows the wave orbit. We expect this to be approximately true given the slack mooring design, but the line tension will cause deviations from the wave orbit, especially when the buoy is displaced farthest from the anchor (Fig.~\ref{fig:mooring}). The line tension is recorded on the GPS buoy, so it should be possible to correct for this effect. This does not appear to change the leading-order behavior but should be evaluated in more detail in future work aiming for a quantitatively accurate correction for the Stokes offset.

One may also worry that the GPS data are insufficiently accurate to estimate the Stokes offset robustly using~\eqref{eqn:stokes}, both because of noise in the measurements and because of insufficient time resolution \parencite[cf.,][]{lenain_contribution_2020}. It should be noted, however, that high-frequency waves contribute less strongly to the Stokes offset than the Stokes drift, given that the offset is proportional to the second rather than third moment of the frequency spectrum \parencite[cf.,][]{webb_wave_2011}. While we are reassured by the leading-order match of the Stokes offset with the residual of the sea surface height budget (Figs.~\ref{fig:heights}c,~\ref{fig:resspec}), an independent evaluation of the accuracy of the GPS-based estimation should be pursued in future work.

\textcite{elipot_measuring_2020} recently proposed to supplement the global sea level measurements from altimetry and tide gauges with GPS measurements from drifting buoys, such as those deployed by NOAA's Global Drifter Program. The Stokes offset should be taken into account in such measurements to the extent that the drifting buoys follow the wave orbit. This is particularly important because substantial trends in wave height---and therefore the Stokes offset---are expected for the 21st century \parencite[e.g.,][]{casas-prat_wind-wave_2024}.

\sloppy

\section*{Acknowledgments}

We thank Shane Elipot, an anonymous second reviewer, and the editor Jerry Smith for helpful comments and suggestions. This research was carried out in part at the Jet Propulsion Laboratory, California Institute of Technology, under a contract with the National Aeronautics and Space Administration (80NM0018D0004). The work was supported in part by NASA grants 80NSSC20K1140 and 80NSSC24K1652 to the California Institute of Technology.

\section*{Data availability statement}

All data can be obtained from NASA's Physical Oceanography Distributed Active Archive Center. The GPS buoy data can be accessed at \url{https://doi.org/10.5067/SWTPR-GPS01}, the mooring CTD data at \url{https://doi.org/10.5067/SWTPR-CTD11}, and the bottom pressure recorder data at \url{https://doi.org/10.5067/SWTPR-BPR01}. The ERA5 data is available from the Copernicus Climate Data Store at \url{https://doi.org/10.24381/cds.adbb2d47}.

\end{document}